\documentclass[12pt,a4paper,final]{iopart}

\usepackage{graphicx}
\usepackage[breaklinks=true,colorlinks=true,linkcolor=blue,urlcolor=blue,citecolor=blue]{hyperref}
\usepackage[section]{placeins}
\usepackage{graphicx}
\usepackage{dcolumn}
\usepackage{bm}

\makeatletter
\@namedef{ver@amsmath.sty}{}
\makeatother
\usepackage{amstext}
\usepackage{color}
\usepackage{float}
\usepackage{bigstrut}
\usepackage[version=3]{mhchem}
\begin{document}
\sloppy
\title[Field-induced transition of the magnetic ground state in \ce{CsCo2Se2}]{Field-induced transition of the magnetic ground state from A-type antiferromagnetic to ferromagnetic order in \ce{CsCo2Se2}}
\author{F. von Rohr$^{1,2}$, A. Krzton-Maziopa$^{3}$, V. Pomjakushin$^{4}$, H. Grundmann$^{1}$, Z. Guguchia$^{1}$, W. Schnick$^{2}$, and A. Schilling$^{1}$}
\address{$^{1}$Department of Physics, University of Zurich, CH-8057 Zurich, Switzerland}
\address{$^{2}$Department of Chemistry, University of Munich (LMU), D-81377 Munich, Germany}
\address{$^{3}$Warsaw University of Technology, Faculty of Chemistry, PL-00-664 Warsaw, Poland}
\address{$^{4}$Lab. for Neutron Scattering, Paul Scherrer Institute, CH-5232 Villigen, Switzerland}
\ead{vonrohr@physik.uzh.ch}
\date{\today}
%
\begin{abstract}
We report on the magnetic properties of \ce{CsCo2Se2} with \ce{ThCr2Si2} structure, which we have characterized through a series of magnetization and neutron diffraction measurements. We find that \ce{CsCo2Se2} undergoes a phase transition to an antiferromagnetically ordered state with a N\'eel temperature of $T_{\rm N} \approx$ 66 K. The nearest neighbour interactions are ferromagnetic as observed by the positive Curie-Weiss temperature of $\Theta \approx$ 51.0 K. We find that the magnetic structure of \ce{CsCo2Se2} consists of ferromagnetic sheets, which are stacked antiferromagnetically along the tetragonal \textit{c}-axis, generally referred to as A-type antiferromagnetic order. The observed magnitude of the ordered magnetic moment at T = 1.5 K is found to be only 0.20(1)$\mu_{\rm Bohr}$/Co. Already in comparably small magnetic fields of $\mu_0 H_{MM}$(5K) $\approx$ 0.3 T, we observe a metamagnetic transition that can be attributed to spin-rearrangements of \ce{CsCo2Se2}, with the moments fully ferromagnetically saturated in a magnetic field of $\mu_0 H_{\rm FM}$(5K) $\approx$  6.4 T. We discuss the entire experimentally deduced magnetic phase diagram for \ce{CsCo2Se2} with respect to its unconventionally weak magnetic coupling. Our study characterizes \ce{CsCo2Se2}, which is chemically and electronically posed closely to the \ce{A_{\textit x}Fe_{2-{\textit y}}Se_2} superconductors, as a host of versatile magnetic interactions.
\end{abstract} 
\maketitle
%
%
\section{Introduction}
Antiferromagnetically ordered (AFM) compounds that can undergo a phase transition to a ferromagnetically ordered (FM) state upon the application of an external magnetic field are referred to as metamagnets \cite{meta1}. If an external magnetic field is large enough, the magnetic moments of all unbound electrons will eventually line up with the applied magnetic field, causing a large overall magnetic moment \cite{meta3}. Commonly, very large magnetic fields are necessary in order to observe so-called spin-flip or spin-flop metamagnetic transitions of compounds with a AFM ground-state (see, e.g., reference \cite{meta1}). \\ \\
\ce{CsCo2Se2} belongs to the layered tetragonal \ce{ThCr2Si2} structure type, which has over 600 intermetallic members, and another 200 intermetallics crystallize in the variant with \ce{CaBe2Ge2} structure \cite{Hoffmann,Johrendt1}. These \ce{AT2X2} structures are the most common crystal structures among ternary compounds. \ce{CsCo2Se2} consists of stacked covalently bonded transition metal-metalloid \ce{Co2Se2} layers, where cobalt is coordinated tetrahedrally in \ce{CoSe4}. The \ce{ThCr2Si2} structure type has recently been found to be a suitable host for exotic physical properties, such as the occurrence of structure-driven quantum critical points at $cT$-$ucT$ phase transitions, e.g. in \ce{SrCo_2(Ge_{1-{\textit x}}P_{\textit x})_2} \cite{Jia,Daigo,Roman,Canfield}, or the heavy-fermionic superconductivity in \ce{KNi2Se2} \cite{McQueen}. Of exceptional scientific interest are the \ce{A_{\textit x}Fe_{2-{\textit y}}Se_2} (A = K, Rb, Cs) superconductors, which also crystallize in this structure type \cite{KFe2Se2,RbFe2Se2,CsFe2Se2}. A close interplay of magnetic order and superconductivity has been discovered in these materials. They exhibit an antiferromagnetic ordering below $T_{\rm N} \approx$ 480 K and become superconducting at $T_{\rm c} \approx$ 30 K \cite{AFe2Se2_bulk1,AFe2Se2_bulk2,AFe2Se2_magn}. The co-existence of these two broken states of symmetry is most likely caused by an intrinsic mesoscopic phase separation, which hosts a complex network of superconducting and antiferromagnetic domains \cite{phase_separation}. The coexistence and competition of magnetic and superconducting phases is a significant common feature of all iron-based superconductors (see, e.g., references \cite{Hosono,anka,Johrendt2,bendele}). Generally, the parent compounds are antiferromagnets, which become superconductors upon hole or electron doping, consequently suppressing the N\'eel temperature \cite{Basov}. \\ \\
Among the compounds \ce{ACo2X2} (with A = K, Rb, Cs, Tl and X = S, Se), which are related to the iron-based superconductors, \ce{CsCo2Se2} and \ce{TlCo2Se2} are the only two antiferromagnets \cite{Greenblatt,Yang13}. The other compounds have been found to order ferromagnetically at temperatures between $T_{\rm C} \approx$ 50 K and 110 K \cite{Greenblatt,Yang13}. In \ce{TlCo2Se2} the magnetic moments were found to order in a non-collinear incommensurate magnetic structure leading to an overall zero net magnetic moment \cite{TlCo2Se2_1,TlCo2Se2_2}. This phase has received considerable experimental and theoretical attention, because it is one of the few cobalt-based compounds with non-collinear magnetic ordering (see, e.g., reference \cite{TlCo2Se2_3,TlCo2Se2_4}). Here, we present the magnetic properties of \ce{CsCo2Se2}, a compound we find to be an A-type antiferromagnet, which displays metamagnetic field-induced transitions initiated in external magnetic fields even below $\mu_0 H < 1$ T.
\section{Experiment}
All samples were prepared by high-temperature solid state synthesis, the sample handling was carried out in an argon or helium glove box under inert atmosphere. Powders of cobalt (99.9 \% purity) and selenium (99.99 \% purity) were thoroughly mixed in a stoichiometric ratio, pressed to a pellet, and placed in a quartz tube. Caesium (99.5 \% purity) was weighted into a quartz container, which was placed into the quartz tube, next to the pellet. The elements were sealed in a double-wall evacuated quartz ampoule and rapidly heated to 1000 $\mathrm{^{\circ}}$C for 2 h. The melt was slowly cooled down to 750 $\mathrm{^{\circ}C}$ at the rate of 6 $\mathrm{^{\circ}C}$/h and then cooled down to room temperature at a rate of 200 $\mathrm{^{\circ}}$C/h. Large dark crystals with a golden lustre were obtained, which could easily be cleaved into plates with flat shiny surfaces.\\ \\
The magnetization was studied using a \textit{Quantum Design} Magnetic Properties Measurements System (MPMS) XL 7 T with a differential superconducting quantum interference device (SQUID) equipped with a reciprocating sample option (RSO). The measurements were performed in a temperature range from $T$ = 5 to 300 K and in fields between $\mu_0 H$ = 0 T and 7 T. The extremely air and moisture sensitive samples were vacuum sealed in quartz ampoules of 5 mm diameter and approximately 10 cm length. In the quartz ampoule the samples were fixed between two quartz cylinders of approximately 5 cm length. Such sample mounting was found to provide a stable surrounding and it produces only a minor background signal to the magnetization measurements \cite{Stephen}. \\ \\
Neutron powder diffraction (NPD) experiments were carried out at the SINQ spallation source at the Paul Scherrer Institute (Switzerland) using the high-resolution diffractometer for thermal neutrons (HRPT) \cite{HRPT}. A wavelength of $\lambda$ = 1.886 \AA \ was employed, measurements were performed at $T=$ 100 K and 1.5 K. The NPD experiments in magnetic fields were carried out with a superconducting magnet (MAO6) that can provide fields up to $\mu_0 H \leq$ 6 T with the magnetic field $H$ vertical to the scattering plane. The sample for the NPD experiments consisted of crushed single crystals loaded into vanadium containers with an indium seal. The refinements were carried out by the Rietveld method using the {\tt FULLPROF} program integrated in the {\tt WINPLOTR} software \cite{Fullprof}. Diffraction maxima were fitted with the Thompson-Cox-Hastings pseudo-Voigt function starting from the instrumental resolution values for the profile parameters U, V, W, and Y. The symmetry analysis was performed using the {\tt ISODISTORT} tool based on the {\tt ISOTROPY} software \cite{isot,isod} and the {\tt BasiRep} program \cite{Fullprof}.
\section{Results and Discussion}
\begin{figure}
\centering
{\includegraphics[width=0.6\textwidth]{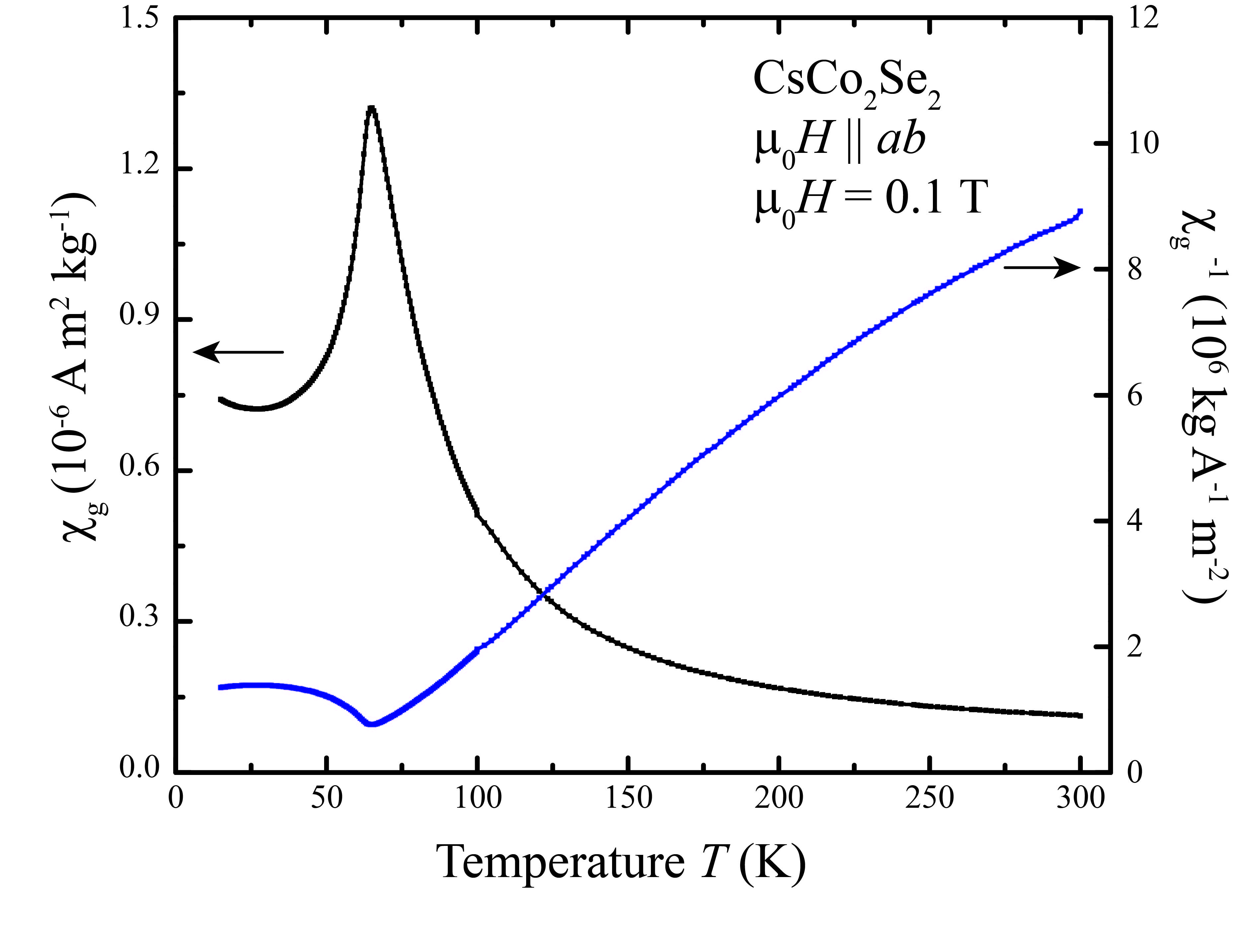}}
\caption{The temperature dependence of the magnetic susceptibility of \ce{CsCo2Se2} measured in a temperature range between 5 K and 300 K in an applied magnetic field of $\mu_{0} H =$ 0.1 T.}
\label{fig:M(T)}
\end{figure}

\begin{figure}
\centering
{\includegraphics[width=0.8\textwidth]{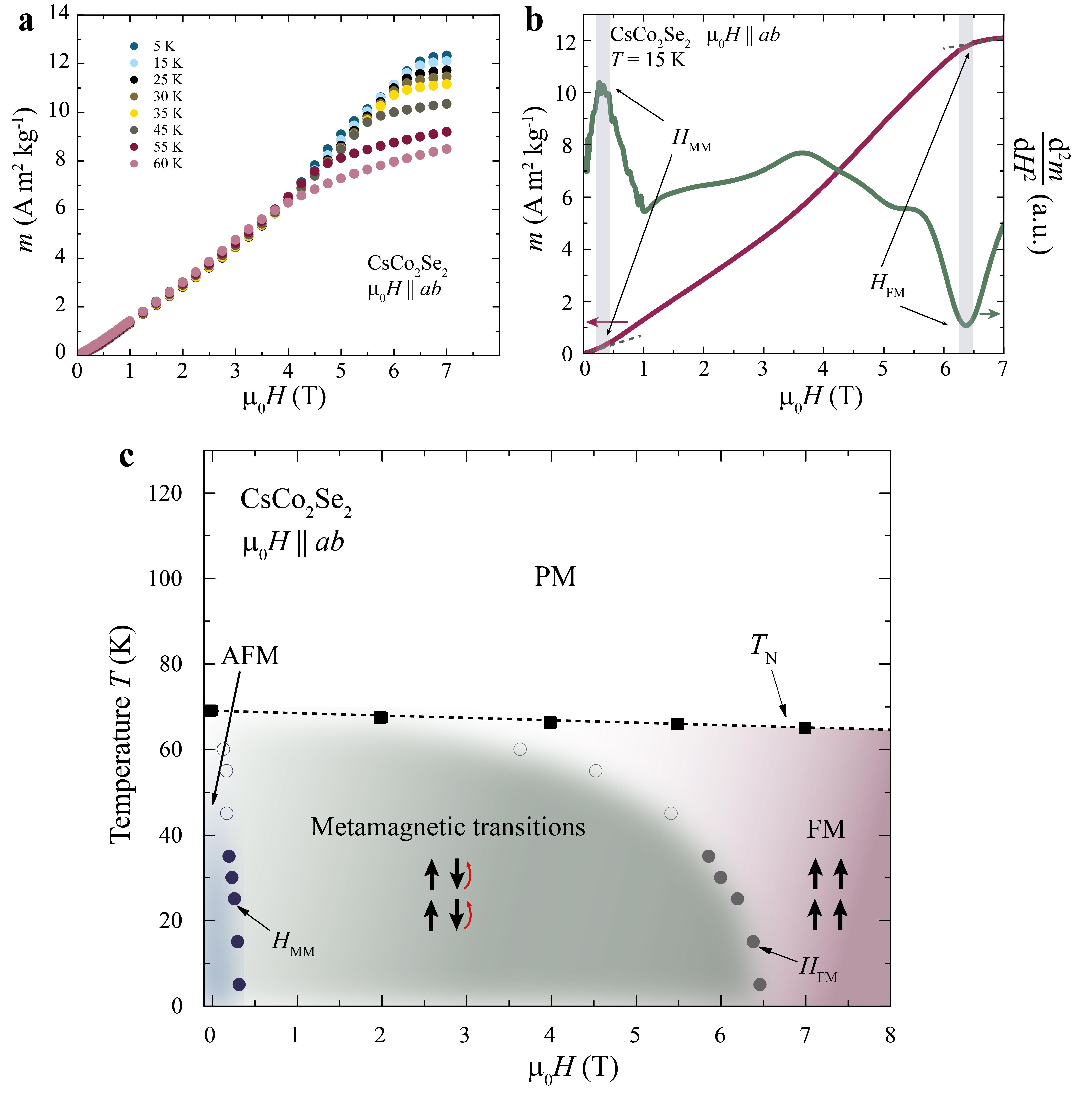}}
\caption{(a) Magnetization $m(H)$ versus the magnetic field $\mu_0 H$ of \ce{CsCo2Se2} for temperatures between 5 K and 60 K (below $T_{\rm N}$). (b) Field dependence of the magnetization at 15 K and its second derivative $d^2m/dH^2$. (c) Phase diagram of \ce{CsCo2Se2} for magnetic fields up to 7 T, determined from field dependent magnetization measurements, according to the criteria illustrated in (b).}
\label{fig:magn_CsCo2Se2}
\end{figure}

In figure \ref{fig:M(T)}, we show the temperature dependent magnetic susceptibility $\chi = M/H$ of single crystals of \ce{CsCo2Se2} in a magnetic field of $\mu_0 H$ = 0.1 T applied perpendicular to the \textit{c}-axis. We find \ce{CsCo2Se2} to be an antiferromagnet (AFM) with a N\'{e}el temperature of $T_{\rm N} \approx$ 66 K, and with a sharp transition, indicating the good quality of the sample. The observed transition temperature is in agreement with the recently reported value for the AFM transition in \ce{CsCo2Se2} by Yang \textit{et al.} \cite{Yang13} It is, however, higher than the previously reported value for polycrystalline samples with a $T_{\rm N} \approx$ 15 K by Huan \textit{et al.} \cite{Greenblatt} Furthermore, the additional transition at 15 K reported in reference \cite{Yang13} is not observed. A transition at these temperatures was only observed for samples which were shortly exposed to air, indicating the formation of a magnetic decomposition product.\\ \\ 
In order to determine the effective magnetic moment $\mu_{\rm eff}$ and the Curie-temperature $\Theta_{\rm CW}$, the Curie-Weiss fit above the AFM transition according to $\chi(T) = \frac{\rm{C}}{T-\Theta_{\rm CW}}$ from $T =$ 300 to 150 K was performed. The effective moment is with a value of $\mu_{\rm eff} \approx$ 1.81 $\mu_{\rm Bohr}$/Co of similar size to comparable intermetallic compounds with the same crystal structure (see, e.g. reference \cite{Jia}). The positive value of $\Theta_{\rm CW} \approx$ 51.0 K indicates that the nearest neighbour interaction between the magnetic moments is ferromagnetic, in contrary to the overall antiferromagnetic ordering.\\ \\
\begin{figure}
\centering
{\includegraphics[width=0.6\textwidth]{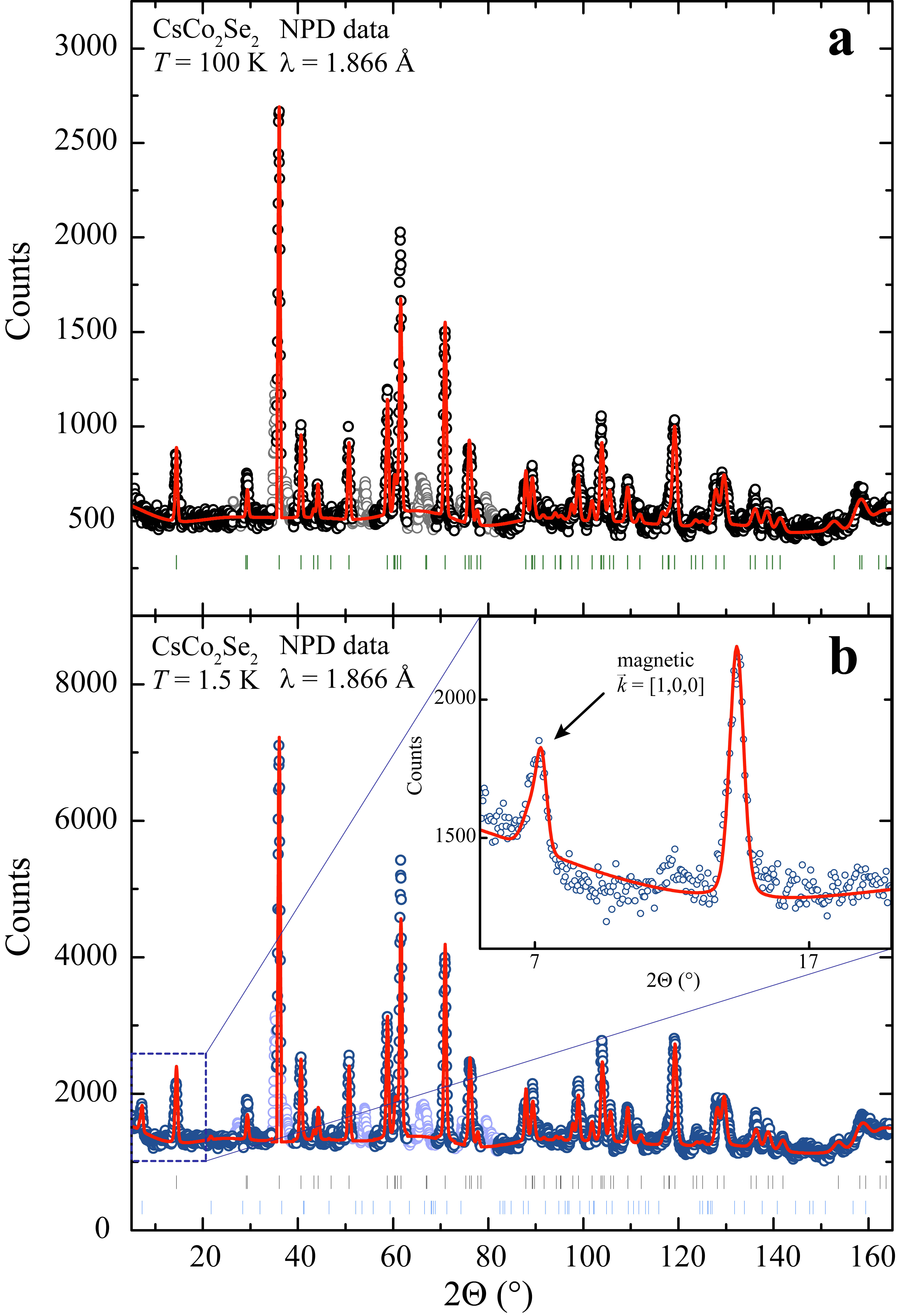}}
\caption{Neutron powder diffraction pattern of polycrystalline \ce{CsCo2Se2} at 100 K (a) and 1.5 K (b) collected with a wavelength of $\lambda = 1.886$ \AA. Black and blue circles, observed patterns; red curves, calculated patterns; black tic marks, calculated peak positions for the crystal structure of \ce{CsCo2Se2}; blue tic marks, calculated peak positions for the magnetic reflections of \ce{CsCo2Se2}.}
\label{fig:NPD_structure}
\end{figure}
In figure \ref{fig:NPD_structure}a, we show the NPD data of pulverized \ce{CsCo2Se2} crystals at 100 K, which was collected with a wavelength of $\lambda = 1.886$ \AA. As expected, most reflections of the diffraction pattern can be well explained with a \ce{ThCr2Si2} structure type model with space group $I4/mmm$. The cell parameters are found to be $a \approx$ 3.842 \AA \ and $c \approx$ 15.041 \AA \ at 100 K. Several additionally observed Bragg reflections cannot be explained solely with this structure model, or with known phases of the Cs-Co-Se phase diagram. They can most likely be attributed to a decomposition product of the extremely air sensitive \ce{CsCo2Se2} compound (see above). A similar sensitivity to moisture and air has earlier been observed for the chemically closely related \ce{A_{1-\textit{x}}Fe_{2-\textit{y}}Se2} phases \cite{Volodja,Stephen}. Furthermore, it was not possible to obtain an improved fit to the structure data with one of the \ce{ThCr2Si2}-related polytypes or with a superlattice structure. A reasonable indexing solution to the extra peaks was obtained with a tetragonal cell with $a \approx 8.815$ \AA \ and $c \approx 9.209$ \AA.
Since there are no obvious candidate impurities for this cell, the additional peaks have not been taken into account for a more accurate magnetic structure refinement (grey points in figure \ref{fig:NPD_structure}). \ce{CsCo2Se2} is an extremely air sensitive compound. It even decomposes in an argon-filled glovebox with almost zero \ce{O2} or \ce{H2O}. The decomposition phases cannot be avoided. This does, however, not affect the scattering experiments or magnetic structure solution presented in the following. \\ \\ 
\begin{figure}
\centering
{\includegraphics[width=0.4\textwidth]{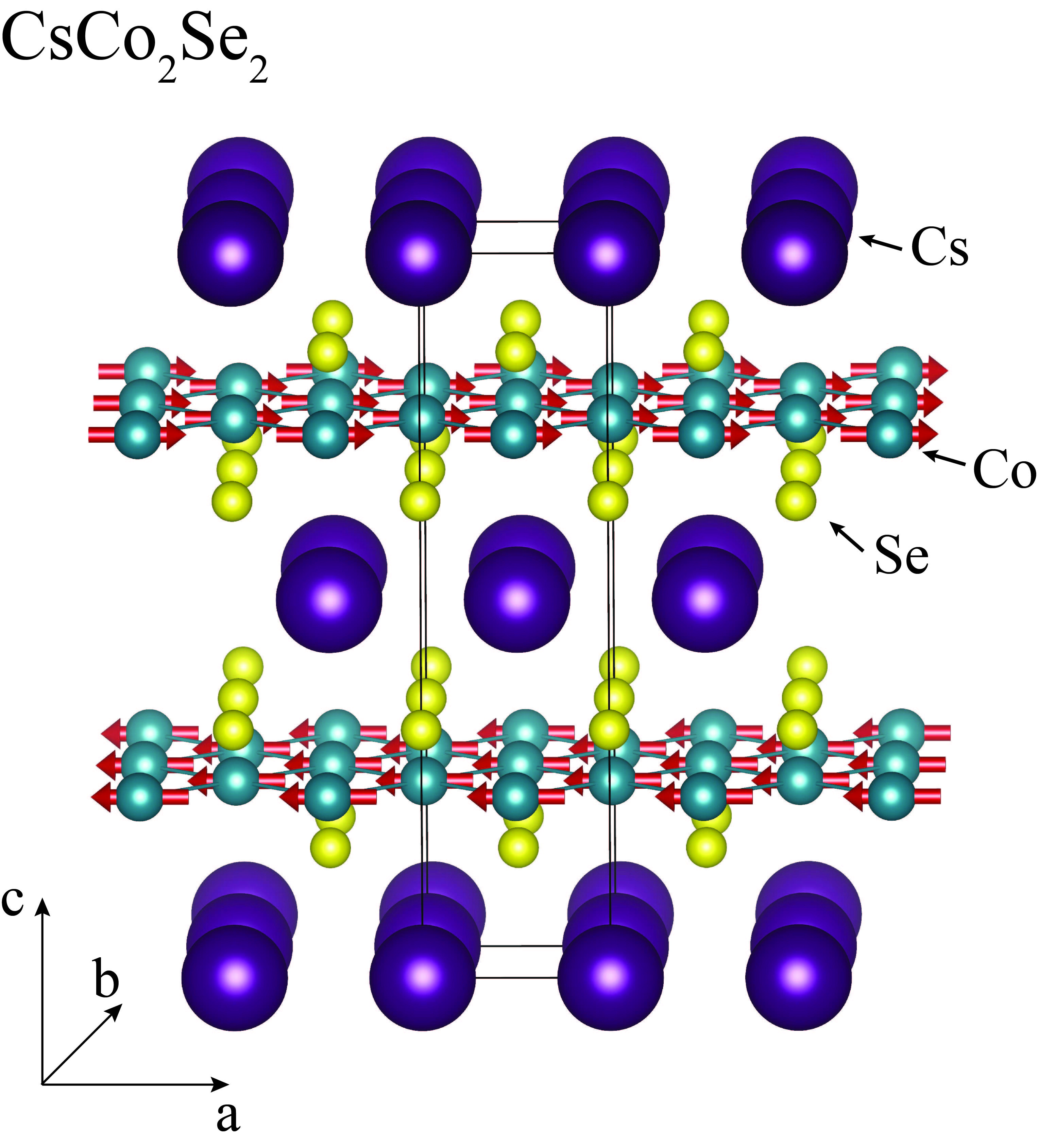}}
\caption{Crystal and magnetic structure of \ce{CsCo2Se2}, the magnetic moments on the cobalt position of the A-type AFM structure are displayed as red arrows.}
\label{fig:magnetic_structure}
\end{figure}
At the bottom of figure \ref{fig:NPD_structure}, we show the NPD data of the same polycrystalline \ce{CsCo2Se2} sample at 1.5 K collected with a wavelength of $\lambda = 1.886$ \AA. We observe a single magnetic diffraction peak at $2\Theta$ = 7.2$^{\circ}$ that corresponds to the (001) reflection of the tetragonal crystal structure on hand. It can be indexed with the propagation vector $\vec{k} = [1,0,0]$. This implies an AFM order for the body centered Bravais lattice. A symmetry treatment was used for the modelling of the magnetic structure, thereby the magnetically ordered structures are described in terms of their magnetic propagation vector and the irreducible representations (see, references \cite{Fullprof,isot,isod, Kovalev}). The decomposition of the magnetic representation for the $I4/mmm$ space group in Kovalev's notation ($\tau_i$ are the allowed irreducible representations of the symmetry groups $G_k$), with the propagation vector $\vec{k} = \vec{k_{15}} = [1,0,0]$ for cobalt in the crystallographic 4d (0,1/2,1/4) position gives the following allowed symmetry solutions: $\tau_2$, $\tau_5$, $\tau_9$, and $\tau_{10}$. The one-dimensional irreducible representations $\tau_2$ and $\tau_5$ give solutions for the magnetic structure, where the magnetic moments are aligned along the crystallographic $c$-axis. However, the magnetic (001) reflection is observed and therefore both these magnetic structures can be rejected. The two-dimensional irreducible representations $\tau_9$ and $\tau_{10}$ give magnetic structure configurations with the magnetic moments aligned along directions in the $ab$ plane. Of these two, the chessboard solution $\tau_9$ can be rejected, because the structure factor $F_{\rm hkl}$ for the Bragg reflection (001) must for $\tau_9$ be zero due to symmetry. Thus, we have a unique solution $\tau_{10}$; here the magnetic moments form FM sheets with the spin direction in the \textit{ab} plane with a magnetic coupling of 0.20(1)$\mu_{\rm Bohr}$/Co. The magnetic structure refinement together with the structural refinement is shown in figure \ref{fig:NPD_structure} and in the inset therein. The corresponding real-space magnetic structure is depicted in figure \ref{fig:magnetic_structure}. It should be noted that the direction of the magnetic moments in the layer cannot be deduced from our experimental data. This solution can also be represented in the Shubnikov magnetic group $P_c2_1/m$ (No. 11.57) with the cobalt atoms in the 4h position $\big(\frac{3}{4}$,$\frac{1}{2}$,$\frac{1}{2}$;$m_x$,0,$m_z\big)$. In this case, the basis transformation from the parent tetragonal paramagnetic group to the monoclinic Shubnikov group is (1,1,0),(0,0,-1),(-1,0,0) with the origin shift $\big(\frac{1}{4}$,$\frac{1}{4}$,$\frac{1}{4}\big)$. This magnetic structure it commonly referred to as A-type antiferromagnetism. \\ \\
\begin{figure}
\centering
{\includegraphics[width=0.5\textwidth]{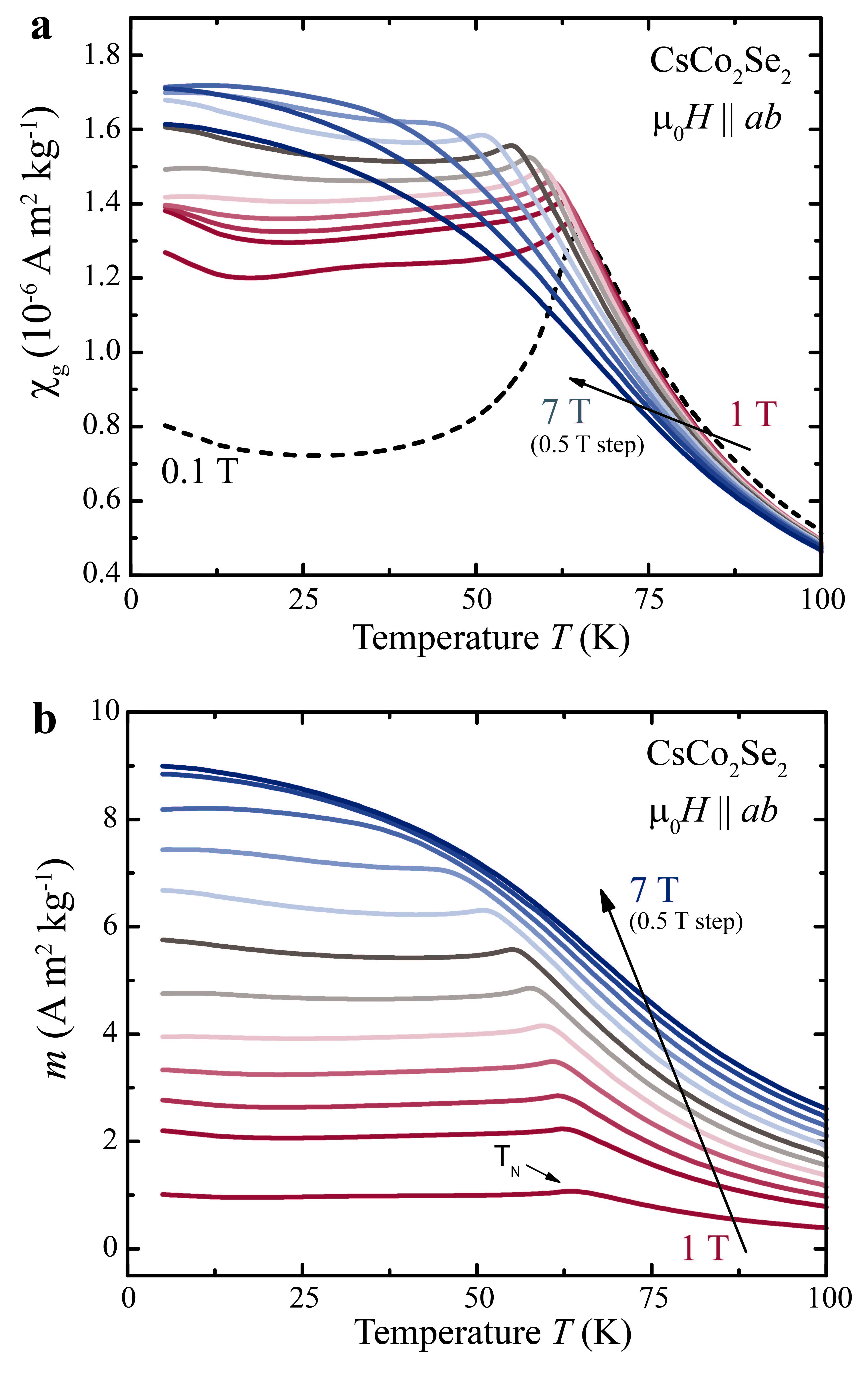}}
\caption{Magnetic susceptibility $\chi$ (a) and the magnetization $m$ (b) in a temperature range of $T$ = 5 K to 100 K in magnetic fields $\mu_{0} H$ from 1 T to 7 T in 0.5 T steps.}
\label{fig:M(T,H)}
\end{figure}
In figure \ref{fig:M(T,H)}, we show the magnetic susceptibility as $\chi = M/H$ (a) and the magnetization (b) in a temperature range between $T$ = 5 K to 100 K with the external field $\mu_{0} H$ perpendicular to the \textit{c} axis of \ce{CsCo2Se2}. These measurements were performed in magnetic fields ranging from $\mu_{0} H =$ 1 T to 7 T in 0.5 T steps. The clearly pronounced metamagnetic transition from a AFM orientation to a FM orientation of the magnetic moments can be observed in these measurements. The transition temperature is shifted only slightly to lower temperatures with higher magnetic fields. A clear saturation of the magnetic moments in a FM or canted AFM alignment is found in fields greater than $\mu_{0} H \approx 6 \ T$, while the transition is observed to be continuous. In figure \ref{fig:magn_CsCo2Se2}a, the field-dependent magnetization of \ce{CsCo2Se2} at 5, 15, 25, 30, 35, 45, 55 and 60 K is shown. As expected, the field dependence of the magnetization of \ce{CsCo2Se2} deviates from a common AFM behaviour and further supports the scenario of a spin reorientation and therefore of a metamagnetic transition. Three distinct regimes can be determined in the field dependent magnetization. The fields necessary for the initialization of the metamagnetic transition is small compared to other metamagnetic materials (see, e.g., \cite{Roy}).\\ \\
Four different magnetic phases can therefore be identified in \ce{CsCo2Se2}: a paramagnetic high-temperature phase (PM), an antiferromagnetically ordered phase (AFM), one or more metamagnetic phase transitions (MM), and a ferromagnetically ordered phase (FM). This nomenclature is thereby chosen on the basis of earlier reports of similar magnetic properties (see, e.g., reference \cite{Zurab}). Here, we have used the deviations from linearity, as observed in the second derivative ($d^2m/dH^2$), in the field-dependent magnetization as a measure for the respective critical fields ($H_{\rm MM}$ and $H_{FM}$). This procedure is illustrated in figure \ref{fig:magn_CsCo2Se2}a for the measurement at $T =$ 15 K. By applying these criteria to the various measured temperatures, we are able to draw a summarizing magnetic phase diagram for \ce{CsCo2Se2} as shown in figure \ref{fig:magn_CsCo2Se2}c.  It should be noted that all of the observed transitions are continuous and that all here determined critical fields are not strict quantities. Furthermore, at higher temperatures the transitions broaden and are less pronounced in the field-dependent magnetization $M(H)$ (represented by the open circles). The observed phase diagram is in qualitative agreement with other metamagnetic materials. Thereby, layered A-type antiferromagnetic materials often undergo metamagnetic transitions in external magnetic fields parallel to the antiferromagnetically ordered spin lattices, because the interlayer AFM coupling is in such an alignment comparably weak. The transitions observed here are in general agreement with earlier observations by Yang J et al. \cite{Yang13}, however in this earlier study a lower $H_{FM}$ was observed. This discrepancy might most likely be connected to a variable off-stoichiometric composition of the compound as it has been extensively studied for the closely-related \ce{A_{1-\textit{x}}Fe_{2-\textit{y}}Se2} and \ce{FeSe} phases (see, e.g. \cite{Stephen,Volodja,FeSe}). \\ \\
\begin{figure}
\centering
{\includegraphics[width=0.6\textwidth]{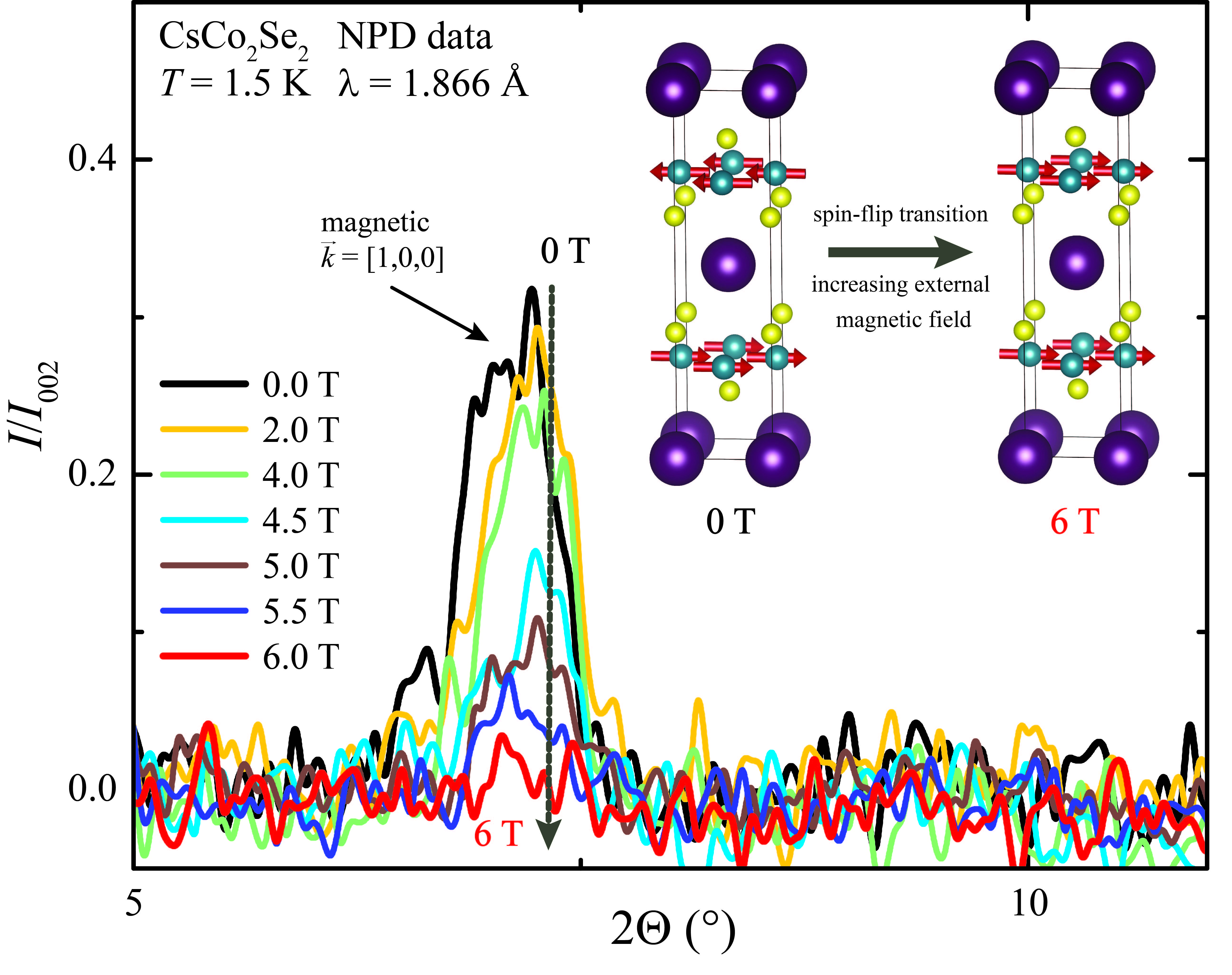}}
\caption{Neutron powder diffraction data of \ce{CsCo2Se2} in magnetic fields $\mu_0 H =$ 0 T, 2 T, 4 T, 4.5 T, 5 T, 5.5 T, and 6 T at $T =$ 1.5 K measured with a wavelength of $\lambda =$ 1.886 \AA. The data is normalized to the intensity of the structural Bragg reflection (002).}
\label{fig:mf}
\end{figure}
In figure \ref{fig:mf}, we show the NPD data of \ce{CsCo2Se2} in magnetic fields $\mu_0 H =$ 0 T, 2 T, 4 T, 4.5 T, 5 T, 5.5 T, and 6 T at $T =$ 1.5 K, measured with a wavelength of $\lambda =$ 1.886 \AA. The data is normalized to the intensity of the structural Bragg reflection (002). For better clarity only the data in the vicinity of the magnetic (001) reflection is shown. The intensity of the magnetic (001) reflection is slightly decreased in magnetic fields of $\mu_0 H =$ 2 T and 4 T. According to the phase diagram (see figure \ref{fig:magn_CsCo2Se2}) in these fields a small magnetic moment in the direction of the external field can be expected due to possible canting where the ferromagnetic moment increases. Therefore, the symmetry of the magnetic order is only slightly perturbed, leading to a corresponding small alternation of the magnetic (001) reflection. In larger magnetic fields, such as $\mu_0 H =$ 4.5 T, 5 T, and 5.5 T, the magnetic (001) reflection is strongly reduced in intensity and in a field of $\mu_0 H =$ 6 T not observable anymore. These findings are in good agreement with the magnetic phase diagram derived from the field-dependent magnetization measurements. A likely scenario for the transition is shown in the inset of figure \ref{fig:mf}, where the A-type AFM structure undergoes a field-induced transition to a FM structure with the magnetic moments laying in the \textit{ab} plane. \\ \\
It should be noted that in the series of the \ce{ACo2X_2} with (A = Cs, Rb, and K and X = Se, S) compounds, the \textit{a}-axis is of similar size for all series members, whereas the \textit{c}-axis increases strongly from \ce{KCo2X2} to \ce{CsCo2X2}. Thereby, the interlayer distance of the \ce{CoX_4}-layers increases and causes the difference in the magnetic behaviour This suggests that a variation of the magnetic properties by chemical variation of the interlayer distance might be of great interest for these and related materials. This is especially expected since the fragile antiferromagnetic order in \ce{CsCo2Se2} can be perturbed in a facile manner by a weak external magnetic field.\\ \\
\section{Conclusion}
In summary, we report on the magnetic properties of \ce{CsCo2Se2}, which we have investigated by a series of NPD and by SQUID magnetometry measurements. We find that \ce{CsCo2Se2} is an antiferromagnet with a N\'eel temperature of $T_{\rm N} \approx$ 66 K with an effective magnetic moment of $\mu_{\rm eff} \approx$ 1.81 $\mu_{\rm Bohr}$/Co. However, its nearest neighbour interactions between the magnetic moments are ferromagnetic. In the collected NPD data, we observe a single magnetic diffraction peak at $2\Theta$ = 7.2$^{\circ}$ below $T_{\rm N}$, which corresponds to the magnetic propagation vector $\vec{k} = [1,0,0]$. We have found a unique solution of the magnetic structure of \ce{CsCo2Se2}, where the magnetic moments are aligned ferromagnetically in the \textit{ab} plane. These FM sheets order antiferromagnetically along the \textit{c}-axis. In external magnetic fields up to $\mu_0 H \geq$ 7 T \ce{CsCo2Se2} undergoes a metamagnetic transition. A spin rearrangement occurs already for a comparably small critical field of $\mu_0 H_{MM}$(5K) $\approx$ 0.3 T with the moments fully ferromagnetically saturated in a magnetic field of $\mu_0 H_{\rm FM}$(5K) $\approx$  6.4 T. Our study characterizes \ce{CsCo2Se2}, which is chemically and electronically posed closely to the \ce{A_{\textit x}Fe_{2-{\textit y}}Se_2} superconductors, as a host of versatile magnetic interactions that likely can be tuned by chemical variation of the interlayer distance. In further studies, the strong correlation between the structure and magnetism in these materials may give new insights into the nature of the magnetic and superconducting interactions in the \ce{ThCr2Si2}-related superconductors and magnets. 
\section{Acknowledgements}
This work was supported by the Swiss National Science Foundation under Grant No. 21-153659. A.K.-M. acknowledges financial support by the National Science Centre of Poland, grant No. DEC-2013/09/B/ST5/03391. The authors thank Stephen Weyeneth, Kazimierz Conder, and Tyrel McQueen for helpful discussions, as well as Christian R\"uegg for his support of the NPD experiments and Denis Sheptyakov for his assistance with the NPD measurements.


\section*{References}
\begin{thebibliography}{99}
	\bibitem{meta1} Stryjewski E and Giordano N 1977 \textit{Advances in Physics} \textbf{26} 487
	\bibitem{meta3} Wohlfarth E P and Rhodes P 1962 \textit{Philosophical Magazine} \textbf{7} 1817
	\bibitem{Hoffmann} Hoffmann R and Zheng C 1985 \textit{J. Phys. Chem.} \textbf{89} 4175
	\bibitem{Johrendt1} Johrendt D, Felser C, Jepsen O, Andersen O, Mewis A, Rouxel J 1997 \textit{Journal of Solid State Chemistry} \textbf{130} 254
	\bibitem{Jia} Jia S, Jiramongkolchai P, Suchomel M, Toby B, Checkelsky J, Ong N, and Cava R J 2011 \textit{Nature Physics} \textbf{7} 207
	\bibitem{Daigo} Hirai D, von Rohr F, and Cava R J 2012 \textit{Phys. Rev. B} \textbf{86} 100505(R)
	\bibitem{Roman} Pobel R, Frankovsky R, Johrendt D 2013 \textit{Zeitschrift f\"ur Naturforschung B} \textbf{68} 581
	\bibitem{Canfield} Torikachvili M S, Bud'ko S L , Ni  N, and Canfield P C 2008 \textit{Phys. Rev. Lett.} \textbf{101} 057006
	\bibitem{McQueen} Neilson J R, Llobet A, Stier A V, Wu L, Wen J J, Tao J, Zhu Y, Tesanovic Z B, Armitage N P, and McQueen T M 2012 \textit{Phys. Rev. B} \textbf{86} 054512
	\bibitem{KFe2Se2} Guo J, Jin S, Wang G, Wang S, Zhu K, Zhou T, He M, and Chen X 2010 \textit{Phys. Rev. B} \textbf{82} 180520(R)
	\bibitem{CsFe2Se2} Krzton-Maziopa A, Shermadini Z, Pomjakushina E, Pomjakushin V, Bendele M, Amato A, Khasanov R, Luetkens H, and Conder K 2011 \textit{J. Phys.: Condens. Matter} \textbf{23} 052203
	\bibitem{RbFe2Se2} Li C H, Shen B, Han F, Zhu X, and Wen H H 2011 \textit{Phys. Rev. B} \textbf{83} 184521
	\bibitem{AFe2Se2_bulk1} Tsurkan V, Deisenhofer J, G\"unther A, Krug von Nidda H A, Widmann S, and Loidl A 2011 \textit{Phys. Rev. B} \textbf{84} 144520
	\bibitem{AFe2Se2_bulk2} Ying J J, Wang X F, Luo X G, Wang A F, Zhang M, Yan Y J, Xiang Z J, Liu R H, Cheng P, Ye G J, and Chen X H 2011 \textit{Phys. Rev. B} \textbf{83} 212502
	\bibitem{AFe2Se2_magn} Pomjakushin V Y, Pomjakushina E V, Krzton-Maziopa A, Conder K, Chernyshov D, Svitlyk V, and Bosak A 2012 \textit{J. Phys.: Condens. Matter} \textbf{24} 435701
	\bibitem{phase_separation} Ricci A, Poccia N, Campi G, Joseph B, Arrighetti G, Barba L, Reynolds M, Burghammer M, Takeya H, Mizuguchi Y, Takano Y, Colapietro M, Saini N L, and Bianconi A 2011 \textit{Phys. Rev. B} \textbf{84} 060511(R)
	\bibitem{Hosono} Kamihara Y, Watanabe T, Hirano M, and Hosono H 2008 \textit{J. Am. Chem. Soc.} \textbf{130} 3296
	\bibitem{anka} Krzton-Maziopa A, Pomjakushina E V, Pomjakushin V Y, von Rohr F, Schilling A, and Conder K 2012 \textit{J. Phys.: Condens. Matter} \textbf{24} 382202
	\bibitem{Johrendt2} Johrendt D 2011 \textit{Journal of Materials Chemistry} \textbf{21} 13726
	\bibitem{bendele} Bendele M, Amato A, Conder K, Elender M, Keller H, Klauss H H, Luetkens H, Pomjakushina E, Raselli A, and Khasanov R 2010 \textit{Phys. Rev. Lett.} \textbf{104} 087003
	\bibitem{Basov} Basov D N and Chubukov A V 2011 \textit{Nature Physics} \textbf{7} 272
	\bibitem{Greenblatt} Huan G and Greenblatt M 1989 \textit{J. Less-Common Met.} \textbf{156} 247
	\bibitem{Yang13} Yang J, Chen B, Wang H, Mao Q, Imai M, Yoshimura K, and Fang M 2013 \textit{Phys. Rev. B} \textbf{88} 064406
	\bibitem{TlCo2Se2_1} Berger R, Fritzsche M, Broddefalk A, Nordblad P, and Malaman B 2002 \textit{J. Alloys Compd.} \textbf{343} 186
	\bibitem{TlCo2Se2_2} Lizarraga R, Ronneteg S, Berger R, Bergman A, Eriksson O, and Nordstr\"om L 2004 \textit{Phys. Rev. B} \textbf{70} 024407
	\bibitem{TlCo2Se2_3} Jeong H K, Valla T, Berger R, Johnson P D, and Smith K E 2007 \textit{EPL} \textbf{77} 27001
	\bibitem{TlCo2Se2_4} Ronneteg S, Felton S, Berger R, and Nordblad P 2003 \textit{J. Magn. Magn. Mater.} \textbf{299}, 53
	\bibitem{Stephen} Weyeneth S, Bendele M, von Rohr F, Dluzewski P, Puzniak R, Krzton-Maziopa A, Bosma S, Guguchia Z, Khasanov R, Shermadini Z, Amato A, Pomjakushina E, Conder K, Schilling A, and Keller H 2012 \textit{Phys. Rev. B} \textbf{86} 134530
	\bibitem{HRPT} Fischer P, Frey G, Koch M, K\"onnecke M, Pomjakushin V, Schefer J, Thut R, Schlumpf N, B\"urge R, Greuter U, Bondt S, and Berruyer E 2000 \textit{Physica B} \textbf{276} 146
	\bibitem{Fullprof} Rodriguez-Carvajal J 1993 \textit{Physica B} \textbf{192} 55
	\bibitem{isot} Stokes H T and Hatch D M 1988 \textit{Isotropy Subgroups of the 230 Crystallographic Space Groups} World Scientific
	\bibitem{isod} Campbell B J, Stokes H T, Tanner D E, and Hatch D M 2006 \textit{J. Appl. Cryst.} \textbf{39} 607
	\bibitem{Volodja} Pomjakushin V Y, Pomjakushina E V, Krzton-Maziopa A, Conder K, and Shermadini Z 2011 \textit{J. Phys.: Condens. Matter} \textbf{23} 156003
	\bibitem{Kovalev} Kovalev O V 1993 \textit{Representations of the Crystallographic Space Groups: Irreducible Representations, Induced Representations and Corepresentations} Gordon and Breach, Amsterdam
	\bibitem{Roy} S.B. Roy, P. Chaddah, V.K. Pecharsky, K.A. Gschneidner Jr. 2008 \textit{Acta Materialia} \textbf{56} 5895
	\bibitem{FeSe} T. M. McQueen, Q. Huang, V. Ksenofontov, C. Felser, Q. Xu, H. Zandbergen, Y. S. Hor, J. Allred, A. J. Williams, D. Qu, J. Checkelsky, N. P. Ong, and R. J. Cava 2009 \textit{Phys. Rev. B} \textbf{79}, 014522
	\bibitem{Zurab} Guguchia Z, Bosma S, Weyeneth S, Shengelaya A, Puzniak R, Bukowski Z, Karpinski J, and Keller H 2011 \textit{Phys. Rev. B} \textbf{84} 144506
\end{thebibliography}
\end{document}